\documentclass[useAMS,usenatbib,usegraphicx]{mn2e}
\def\be{\begin{equation}}
\def\ee{\end{equation}}
\def\bea{\begin{eqnarray}}
\def\eea{\end{eqnarray}}

\def\nn{\nonumber}

\def\dm{\Delta {\rm M}}
\def\ms{{\rm M_{\odot}}}

\def\bo{B_{o}}

\def\rnin{R_{9}}

\def\la{ \langle}

\def\ga{\gamma}

\def\la{\lambda}

\def\rmo{{\rm R_{M0}}}

\def\mdot{\ifmmode \dot M \else $\dot M$\fi}    
\def\mxd{\ifmmode \dot {M}_{x} \else $\dot {M}_{x}$\fi}
\def\med{\ifmmode \dot {M}_{Edd} \else $\dot {M}_{Edd}$\fi}
\def\bff{\ifmmode B_{{\rm f}} \else $B_{{\rm f}}$\fi}

\def\apj{\ifmmode ApJ \else ApJ \fi}    
\def\apjl{\ifmmode  ApJ \else ApJ \fi}    %
\def\apjs{\ifmmode  ApJS \else ApJS \fi}
\def\aap{\ifmmode A\&A \else A\&A\fi}
\def\aaps{\ifmmode A\&AS \else A\&AS\fi}    %
\def\mnras{\ifmmode MNRAS \else MNRAS \fi}    %
\def\nat{\ifmmode Nature \else Nature \fi}
\def\prl{\ifmmode Phys. Rev. Lett. \else Phys. Rev. Lett.\fi}
\def\prd{\ifmmode Phys. Rev. D. \else Phys. Rev. D.\fi}
\def\pasp{\ifmmode  PASP \else PASP \fi}

\def\ms{\ifmmode M_{\odot} \else $M_{\odot}$\fi}    
\def\na{\ifmmode \nu_{A} \else $\nu_{A}$\fi}    
\def\nk{\ifmmode \nu_{K} \else $\nu_{K}$\fi}    
\def\ns{\ifmmode \nu_{{\rm s}} \else $\nu_{{\rm s}}$\fi}
\def\no{\ifmmode \nu_{1} \else $\nu_{1}$\fi}    
\def\nt{\ifmmode \nu_{2} \else $\nu_{2}$\fi}    
\def\ntk{\ifmmode \nu_{2k} \else $\nu_{2k}$\fi}    
\def\dnmax{\ifmmode \Delta \nu_{max} \else $\Delta \nu_{2max}$\fi}
\def\ntmax{\ifmmode \nu_{2max} \else $\nu_{2max}$\fi}    
\def\nomax{\ifmmode \nu_{1max} \else $\nu_{1max}$\fi}    
\def\nh{\ifmmode \nu_{\rm HBO} \else $\nu_{\rm HBO}$\fi}    
\def\nqpo{\ifmmode \nu_{QPO} \else $\nu_{QPO}$\fi}    
\def\nz{\ifmmode \nu_{o} \else $\nu_{o}$\fi}    
\def\nht{\ifmmode \nu_{H2} \else $\nu_{H2}$\fi}    
\def\ns{\ifmmode \nu_{s} \else $\nu_{s}$\fi}    
\def\nb{\ifmmode \nu_{{\rm burst}} \else $\nu_{{\rm burst}}$\fi}
\def\nkm{\ifmmode \nu_{km} \else $\nu_{km}$\fi}    
\def\ka{\ifmmode \kappa \else \kappa\fi}    
\def\dn{\ifmmode \Delta\nu \else \Delta\nu\fi}
\def\mdotsix{\ifmmode\dot{M}_{16} \else \dot{M}_{16}\fi}

\def\rhof{\ifmmode \rho_{5} \else \rho_{5}\fi}
\def\rhos{\ifmmode \rho_{6} \else \rho_{6}\fi}
\def\mdotcr{\ifmmode \dot{M}_{cr} \else  \dot{M}_{cr}\fi}
\def\tohm{\ifmmode t_{ohmic} \else  t_{ohmic} \fi}

\def\tdif{\ifmmode t_{diff} \else  t_{diff} \fi}
\def\tacc{\ifmmode t_{accr} \else  t_{accr} \fi}
\newcommand{\gsimeq}{\mbox{$\, \stackrel{\scriptstyle >}{\scriptstyle
\sim}\,$}}
\newcommand{\lsimeq}{\mbox{$\, \stackrel{\scriptstyle <}{\scriptstyle
\sim}\,$}}

\title[Field restructuring in accreting white dwarfs]{Is there evidence for
field restructuring or decay in accreting magnetic white dwarfs?} \vskip 3cm

\author[C.M. Zhang, D. T. Wickramasinghe and L. Ferrario]
{C.M. Zhang$^{1}$
 \thanks{E-mail: zhangcm@bao.ac.cn,
dayal.wickramasinghe@anu.edu.au, lilia@maths.anu.edu.au}
  , D. T.  Wickramasinghe$^2$, and Lilia Ferrario$^2$\\
$^{1}$   National Astronomical Observatories,
 Chinese Academy of Sciences, Beijing 100012, China\\
$^{2}$Mathematical Sciences Institute, The Australian National
University, Canberra ACT 0200, Australia}

\begin{document}
\date{Received date ; accepted date}

\maketitle
\begin{abstract}

The evolution of the magnetic field of an accreting magnetic white
dwarf with an initially dipolar field at the surface has been
studied for non-spherical accretion under simplifying assumptions.
Accretion on to the polar regions tends to advect the field toward
the stellar equator which is then buried. This tendency is countered
by Ohmic diffusion and magneto-hydrodynamic instabilities.  It is
argued that if matter is accreted at a rate of $\dot{M}_{\rm crit}
\sim 10^{16}$ g~s$^{-1}$ and the total mass accreted exceeds a
critical value $\Delta M_{\rm crit} \sim 0.1-0.2\ms$, {\bf the field may be expected
to be restructured, and the  polar
field  to be reduced} reaching a minimum value of
$\sim 10^3$~G (the ``bottom field'') independently of the initial
field strength. Below this critical accretion rate, the field
diffuses faster than it can be advected, and accretion has little
effect on field strength and structure.

In polars, where the magnetic field strength ($\sim 10^7 - 10^8$~G)
is strong enough to lock the magnetic white dwarf into synchronous
rotation with the orbit and a disc does not form, magnetic braking
is severely curtailed as the stellar wind from the secondary becomes
trapped in the combined magnetosphere of the two stars. The mass
transfer rate in such systems is typically $\lsimeq 10^{16}$
g~s$^{-1}$, and field restructuring  is not
expected. In systems with fields not strong enough to achieve
synchronism and where accretion occurs via a truncated disc (the
intermediate polars), normal magnetic braking may be expected. The
mass transfer rates are then typically $\gsimeq 10^{16}$ g~s$^{-1}$
above the $2-3$ hour Cataclysmic Variable period gap, and thus
{\bf a significant reduction of the polar  field strength} could occur if such a system
accumulates the required critical mass $M_{\rm crit}$. However, due
to mass loss during nova eruptions, only a small fraction of such
systems (those that first come into contact at long orbital periods)
may accumulate sufficient mass to reach the bottom field
configuration.  We argue that the observed properties of the
Magnetic Cataclysmic Variables (MCVs) can generally be explained by a model
where the field is at most only partially {\bf restructured} due to accretion.
 If there are systems that have reached the bottom field, they may be found
among the dwarf novae, and be expected to exhibit quasi periodic
oscillations.

\end{abstract}

\begin{keywords}
stars: novae, cataclysmic variables; stars: dwarf novae; stars:
binaries: close; stars: white dwarfs; stars: magnetic fields.
\end{keywords}


\section{Introduction}

In the Magnetic Cataclysmic Variables (MCVs) which comprise some
$25$\% of the Cataclysmic Variables (CVs), there is direct or
indirect evidence for a magnetic field that influences the accretion
flow on to the white dwarf. MCVs {\bf tend to } fall into two
distinct classes: the AM Herculis-type variables (or polars) and the
intermediate polars (IPs) (see Warner 1995 for a comprehensive
review; Patterson 1994; Wood et al. 2000; Wickramasinghe and
Ferrario 2000, hereafter WF00).

In the vast majority of the polars, the magnetic field of the white
dwarf is strong enough to magnetically lock it into synchronous
rotation with the orbital period through magnetic interaction with the
secondary star. Thus, in polars the spin period $P_{\rm s}$ equals the
orbital period $P_{\rm orb}$. Only a handful of polars are slightly
asynchronous ($P_s/P_{\rm orb} \sim 0.99$), likely due to recent nova
events during which synchronism was broken. In polars, accretion
occurs directly on to the magnetic white dwarf (MWD) without the
formation of an accretion disc. The magnetic fields in polars are well
determined through the observation of cyclotron and Zeeman features,
and are found to be in the range $7 - 230$~MG (WF00).

\begin{figure}
\includegraphics[width=8.0cm, angle=0]{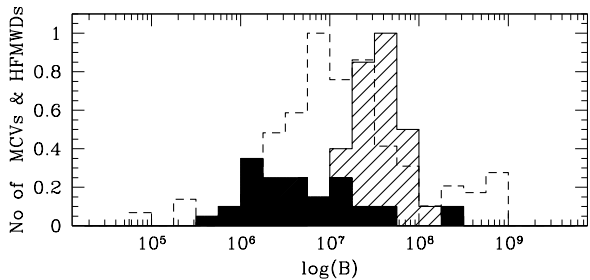}
\caption{Magnetic field distribution of the polars (solid histogram),
  and the intermediate polars (shaded histogram). The IP magnetic
  fields are calculated by assuming that these systems are at their
  equilibrium spin period (Norton et al. 2004), so the
  fields are not measured directly like in polars. The dashed
  histogram shows the distribution of fields in the high field
  isolated magnetic white dwarfs.}
\label{magCVs_fields}
\end{figure}

The IPs are characterised by the presence of asynchronously rotating
white dwarfs with $P_s/P_{\rm orb} \sim 0.01 - 0.9$. In these
systems, accretion occurs via a truncated accretion disc and
magnetically confined accretion curtains (e.g. Ferrario \&
Wickramasinghe 1993; Ferrario, Wickramasinghe \& King 1993).  A
direct field measurement is only available for one IP: V405 Aurigae
for which low-resolution circular spectropolarimetry has revealed
the presence of cyclotron harmonics corresponding to a field of
31.5~MG (Piirola et al. 2008) which is in the range of fields found
in the polars. However, as a class, IPs tend to be at the lower end
($\sim 5-20$~MG) of the field distribution in polars, as it is
inferred by the circular polarization survey of IPs conducted by
Butters et al. (2009). Fields estimated on the assumption that the
white dwarf is in spin equilibrium with an accretion disc generally
confirms this expectation (e.g. Norton, Wynn \& Somerscales 2004).
The field distributions of the polar and the IPs with fields
determined either directly (in the case of polars), or with the
assumption of spin equilibrium and/or broadband circular
polarization measurements (in the case of IPs) are shown in Figure
\ref{magCVs_fields}.

There are some parallels and differences with the observed properties
of the isolated Magnetic White Dwarfs (MWDs) which fall into two
disjoint groups: the high field magnetic white dwarfs (HFMWDs) with
fields in the range $10^6 - 10^9$~G, and the low field magnetic white
dwarfs (LFMWDs) with fields $< 10^5$~G (see WF00 and Wickramasinghe \&
Ferrario 2009 for a comprehensive review).  The HFMWDs comprise some
$12-15$\% of all WDs detected in volume-limited surveys, while the
LFMWDs form a group that is at least as numerous. The field
distribution of the isolated HFMWDs is compared with that of the MCVs
and is also shown in Figure \ref{magCVs_fields}. Although it is this
group that is most likely to be associated with the MCVs, we note that
there are distinct differences in the two distributions. The peak of
the distribution for isolated HFMWDs occurs at $\sim \times 10^7$~G
compared to $3\times 10^7$ for the MCVs. Furthermore, very high
magnetic field systems ($10^8-10^9$)are not as well represented among
the MCVs.

The magnetic properties of the isolated LFMWDs are less well
established due to current limitations of measuring fields $\lsimeq
10^3$~G. It is possible that all WDs are magnetic at some level and
thus belong to the LFMWD group or it could be that most stars are
essentially non magnetic. The LFMWDs are likely to be associated
with the vast majority of the CVs that shows no direct evidence for
coherent pulsations at the spin period of the white dwarf as
expected if there is no magnetically channeled accretion on to the
surface of the white dwarf. On the other hand, the study of
Quasi-Periodic Oscillations (QPO) has indicated the possible
presence of weak
magnetic fields in some dwarf novae (e.g. Warner \& Woudt 2002,
2008; Warner \& Pretorius 2008).

The differences between the observed field distributions of the HFMWDs
and MCVs may be due to observational selection effects or be caused by
accretion which affects the surface magnetic field of the white
dwarf. Isolated HFMWDs show no evidence of field decay on the time
scale of the age of the galactic disc and these conclusions are
generally consistent with theoretical estimates of time scales for
free Ohmic decay of the dipolar mode. On the other hand, Cumming
(2002, 2003, 2004) calculated the Ohmic decay modes for spherically
accreting white dwarfs with a liquid interior allowing for
compressional heating, and found Ohmic decay time scales of several
billion years, comparable to the orbital evolutionary time scales of
typical CVs. These calculations appear to suggest that field decay may
play a role in our understanding of the observed properties of MCVs.
However, it is unclear whether the assumption of spherical accretion
is applicable to polars and IPs where accretion occurs via funnels
onto magnetic polar caps.

In this paper, we ask the question whether the observations of IPs
and Polars can be used to distinguish between the following three
possibilities; (i) accretion induces a {\bf a net decrease in polar field strength and changes
 the field structure}, (ii)
accretion screens the field during phases of accretion but the field
subsequently re-emerges to almost the original value when accretion
ceases, or (iii) that accretion has little or no effect on the field
strength and structure.  Our study is based on a model of
non-spherical accretion on to an initially dipolar magnetic field
which was first developed for accreting neutron stars. Here, we
extend these calculations to conditions appropriate to accreting
white dwarfs. The model predicts that the magnetic polar cap widens
as material is accreted with the field being advected towards the
equatorial regions by the ensuing hydromagnetic flow (e.g., Romani
1990; Konar \& Choudhury 2004; Payne \& Melatos 2004; Zhang \&
Kojima 2006, hereafter ZK06).  In the idealised case where Ohmic
diffusion or MHD instabilities are assumed unimportant, the polar
field reaches an asymptotic value (the so-called ``bottom field'')
as the equatorial field becomes fully submerged by the advected
matter.  One of the interpretations of the observed properties of
neutron stars in the Low Mass X-ray Binaries (LMXBs) and the binary
millisecond pulsars (BMSPs) is that such a bottom field has indeed
been reached at a value of $\sim 10^{8}$ G (Wijnands \& van der Klis
1998).

The paper is organized as follows. We describe the basic model in
section \ref{basic_model}, where we present and discuss the
equations and solutions for the evolution of the polar field of the
white dwarf in the case where diffusive effects or MHD instabilities
are considered unimportant.  The expected time scales for Ohmic
diffusion and advection for conditions appropriate to accreting
white dwarfs are discussed in section \ref{ohmic_diffusion}. The
application to the MCVs is presented in section \ref{app_MCVs} where
we discuss the various accretion regimes and the possible evidence
for field restructuring and {\bf decline}.  Finally, our main
conclusions are summarized in section \ref{conclusions}.

\section{The basic field advection model}

\subsection{Advection of polar magnetic zone}\label{basic_model}


ZK06 developed a simple analytical model which encapsulates the basic
physics of non-spherical accretion onto a magnetised neutron star with
a centred dipolar field. On this model, as matter accretes along
dipolar field lines onto the polar cap regions, the accreted matter
moves both downwards into the deeper layers of the star, and in a
perpendicular direction dragging the field lines with the flow towards
the equatorial regions.  The net effect is an increase in the width of
the polar cap region which yields a decrease in the polar field
strength $B$.

Following ZK06, we assume that as accretion proceeds, and the polar
field $B$ decreases, the magnetospheric radius $R_{\rm M}$ continues
to be given by the standard expression (Ghosh \& Lamb 1979) for a
dipolar magnetic field \be R_{\rm M} = \phi R_{\rm A} = 1.2\times
10^{11} ({\rm cm})\phi \dot{M}^{-2/7}_{\rm 16} B_7^{4/7}R_9^{12/7}
m^{-1/7}\,\,\, \label{ra} \ee where $R_{\rm A}$ is the spherical
Alfven radius, and $\phi$ is a parameter which is estimated to be
$\sim 0.5$ for disc accretion, but is model dependent (e.g. Li \&
Wang 1999). Further, $m = M/M_{\odot}$ is the white dwarf mass $M$
in units of solar masses, $R_9$ is the white dwarf radius in units
of $10^9$~cm, $\dot{M}_{16}$ and $B _{7}$ are the accretion rate in
units of $10^{16}$ g~s$^{-1}$ and field in units of $10^{7}$ G,
respectively. The bottom field $B_{\rm f}$ is reached when the
magnetospheric radius $R_{\rm M}=R$, that is,

\be \bff = 2.8 \times 10^3\, (G)\, \mdotsix^{1/2} m^{1/4}R^{-5/4}_9
\phi^{-7/4}\,. \label{bmin} \ee

In the ZK06 model, the approach to the bottom field state is described
by assuming that the motion that drags and restructures the field
occurs in the regions of the star that are built up by accreted
matter.
It is further assumed that when the mass of this layer exceeds a
critical value $\Delta M_{\rm crit}$, the accreted matter becomes
incorporated
into the interior of the star, and does not partake in further
restructuring of the field.  These assumptions, combined with the
equations of continuity and magnetic flux conservation, lead to the
following equation for the polar field $B$ in terms of the
initial polar field  $\bo$ before accretion,  and the amount
of mass $\Delta M$ accreted
\be
\label{bt} B = \frac{\bff}{\{1 -
  \left[C/\exp(y)-1\right]^2\}^{7/4}}\,,
\ee where $y=\displaystyle{\frac{2\xi\Delta M}{7\Delta M_{\rm
crit}}}$, $\dm=\mdot t$, $C = 1+\sqrt{1-x_0}$ and $ x_0 =
\left(\displaystyle{\frac{\bff}{\bo}}\right)^{4/7} =
\displaystyle{\frac{R}{\rmo}}$, where $\rmo$ is the initial
magnetospheric radius. The parameter $\xi$ allows for slippage of
field lines across matter with $\xi=1$ corresponding to the ideal
case of flux freezing.

Following ZK06, and by analogy with neutron stars, we assume that
$\Delta M_{\rm crit}$ is determined by the condition that the
density at the base of the accreted layer equals the mean density of
the white dwarf $\rho_{\rm m}=2\times 10^{5}
(M/0.6\ms)R_9^{-3}$~(g~s$^{-1})$.  Hydrostatic balance gives the
pressure beneath a layer of mass {\bf $\Delta M$} to be

\bea
P & = &\frac{\dm~ g}{4\pi R^2 } \\ \nn
       & = & 2.1\times 10^{21} (\mbox{erg~cm$^{-3}$}) \left(\frac
{\dm}{0.1\ms}\right)
 {}\left(\frac{M}{\ms}\right) \rnin^{-4}\;,
\eea where $g=GM/R^{2}$ is the acceleration due to gravity and it is
assumed that the pressure is that of a degenerate electron gas with
an equation of state (Shapiro \& Teukolsky 1983) \be P =K\rho^{5/3}=
6.8\times 10^{20}~ (\mbox{erg~cm$^{-3}$})\rho_5^{5/3}\;,  \nn \ee
{\bf where $\rhof=\rho/10^{5} (g/cm^{3})$}.  The mass density
beneath this layer is \bea
\rho & = & \left(\frac{\dm~g}{4\pi R^{2} {\rm K}}\right)^{3/5} \\
\nn & = & 2\times 10^{5} \left(\mbox{erg~cm$^{-3}$}\right)
\left(\frac{\dm}{0.1\ms}\right)^{3/5}\,\left(\frac{M}{\ms}\right)^
{3/5} \rnin^{-12/5} \eea and its thickness is \bea \label{thickness}
H & = & \left(\frac{\dm}{4\pi
  R^{2}}\right)^{2/5}\,\mbox{K}~g^{-3/5}\\\nn
  & = & 0.85\times10^{8} (\mbox{cm}) \left(\frac{\dm}
  {0.1\ms}\right)^{2/5} \,\left(\frac{M}{\ms}\right)^{-3/5} \rnin^
{2/5}
\eea
The average density of a white dwarf of mass $M=0.6\ms$ is $\rho =2\times
10^5$~g~cm$^{-3}$ so that with our assumption we have $\Delta M_{\rm
  crit} \sim 0.1 \ms$. This attained at a depth $H/R \sim 0.1$ below
the white dwarf surface.

\subsection{Advection vs Ohmic diffusion}\label{ohmic_diffusion}

In the previous section, we discussed non-spherical accretion in the
advective limit with the parameter $\xi$ allowing for the possibility
of inefficient coupling of matter on to field lines in a global sense.
However, we expect that Ohmic diffusion and MHD instabilities will act
to counter the advection of the field differently in different regions
of the star (e.g. equatorial and polar) so that the use of a global
parameter $\xi$ is too simplistic.

Generally, MHD magnetic field evolution is described by the induction
equation which includes the contributions of two primary terms, MHD
flow and magnetic diffusivity. The relative importance of these terms
can be measured by the magnetic Reynolds number $\cal{R}_{\rm em}$
which
is the ratio of the Ohmic diffusion time to the flow characteristic
time which can be estimated from
\be
{\cal R}_{\rm em}=\frac{t_{\rm diff}}{t_{\rm acc}} = \frac{VL}{\eta}
\ee
where $V$ is the flow velocity, $L$ is the length-scale over which the
field changes, $\eta=c^2/4\pi\sigma$ is the magnetic diffusivity,
and $\sigma$ the electrical conductivity. In our case, we can take
the flow velocity to be the velocity with which the radius of the star
shrinks as the mass of the star increases as a consequence of the
mass-radius relationship of white dwarfs (e.g. Shapiro \& Teukolsky
1983),
 \be
v_{\rm acc} = \frac{\mdot}{4\pi R^{2}\rho} = 8\times 10^{-10} (\mbox
{cm~s$^{-1}$}) \mdot_{\rm 16}\rnin^{-2}\rhos^{-1}\;, \ee {\bf where
$\rhos=\rho/10^{6} (g/cm^{3})$}. Cumming (2002) has argued that
accreting WDs should have liquid interiors, with the electrical
conductivity set by collisions between the degenerate electrons and
the non-degenerate ions. For a non-relativistic degenerate gas, and
for a composition appropriate to CO white dwarfs, approximately
(e.g.~Yakovlev \& Urpin 1980; Itoh et al.~1983; Schatz et al.~1999),
$\sigma = 6.4 \times 10^{20}\ ( {\rm
  s^{-1}}) \rhos$. The magnetic Reynolds number then becomes
\be {\cal R}_{\rm em}=\frac{\sigma \mdot L}{c^{2}R^2 \rho} = {\bf
7.1 }\mdot_{\rm 16}\rnin^{-1}\frac{L}{R}\;.
 \ee For the polar field,
diffusion occurs laterally across the surface, so an appropriate
length scale is $L\sim R$.  For ${\cal R}_{\rm em} \gg 1$ (${\cal
R}_{\rm em} \ll 1 $), the accretion flow (Ohmic diffusion) will
dominate the field evolution process.  Thus, for accretion rates
above a critical rate of about \be \mdot_{\rm br}(p) =
1.4\times10^{15}{\rm (g~s}^{-1}) \rnin  \;, \ee advection will
dominate over diffusion and the bottom-field state will be reached
when {\bf a critical mass mass $\Delta M_{\rm crit}/\xi$ is
accreted}. For significantly lower accretion rates ($\mdot \ll
\mdot_{\rm br}(p)$), the field will diffuse back towards the pole
faster than it will be advected towards the equator, and accretion
will have a minimal effect on the polar field strength.

On the other hand, for the equatorial field it would be appropriate
to take a length scale $L=H$, the thickness of the region that
partakes in the dragging of field lines (the electric current zone).
With $H\sim 0.1 R$, we find that the equatorial field will be
effectively buried above a critical accretion rate of $\mdot_{\rm
br}(eq) = 1.4\times10^{16} ({\rm g~s}^{-1})$ or, more generally,
setting $L=H$ and using equation \ref{thickness},

\be
\mdot_{\rm br}(eq) = 2.4\times10^{16} (\mbox{g~s}^{-1}) R_9^{7/5}
\left(\frac{M}{\ms}\right)^{3/5} \left(\frac{\Delta
  M}{0.1\ms}\right)^{-2/5}
\ee

Above this rate, we expect a bottom field to be reached when a
critical mass $\Delta M_{\rm crit}/\xi$ is accreted. Somewhat below
this rate, the field will diffuse outwards faster than it will sink
into the equatorial region, and we may expect the field to be only
temporarily enhanced in the equatorial zone.

It is also instructive to consider separately the time scales for
accretion and Ohmic decay. The former is given by
\bea \label{eq:tacc2}
t_{\rm acc} & = & \frac{\dm}{\mdot} \\ \nn
& = & 6.3\times 10^{8} (\mbox{yr}) \left(\frac{\dm}{0.1\ms}\right)
~\left(\frac{\mdot}{10^{16} \mbox{g~s$^{-1}$}}\right)\; .
\eea
If we consider diffusion out of a region of length scale $L$, we
obtain
$$t_{\rm diff}={\cal R}_{\rm em}t_{\rm acc}=3\times 10^{10} (\mbox
{yr})
\left(\frac{L}{R}\right)^2\rho_5R_9^2$$ For $L\sim R$, the Ohmic
decay time scale of the dipolar component of the field is $t_{\rm
diff} (p) \sim 3\times 10^{10}$~yr.  On the other hand, for the
equatorial field, $L\sim 0.1 R$ and the decay time scale is $\sim
3\times 10^8$~yr. More generally, the Ohmic diffusion time-scale
across an accreted mass $\Delta M$ in the equatorial zone is

\bea \label{eq:tdif2}
t_{\rm diff}(eq) &=&\! 4\pi\sigma H^2/c^2\\
&=&\! 4.4 \times 10^{8} (\mbox{yr})
\left(\frac{\dm}{0.1\ms}\right)^{7/5}\!\!\!
\left(\frac{M}{\ms}\right)^{-3/5}\!\!\! \rnin^{-8/5} .\nn \eea

\section {Application to Magnetic Cataclysmic Variables}\label
{app_MCVs}

In the strongly advective limit - that is at high enough accretion
rates where Ohmic diffusion cannot counter the advection of the field,
we may expect the properties of accreting white dwarfs to fall into
the following broad regimes.  For these estimates, we assume that
initially, the field is strong enough to satisfy the condition $x_0\ll
1$ (or $R\ll R_{\rm M0}$), and also set $\xi=1$.

\begin{itemize}

\item {\bf Low values of the accreted mass ($\mathbf{\dm <  0.001
\ms}$)}

If $\dm<\Delta M_b \ll \Delta M_{crit} $ (with $x_0\ll 1$) ,the
solution to equation
(\ref{bt}) is given approximately by
\be
B = \frac{\bo}{(1 +\displaystyle{\frac{4\dm}{7\Delta M_b}})^{7/4}}
\,, \label{bdm1}
\ee
where $\Delta M_b = 0.5x_0\Delta M_{crit}= 0.5(R/\rmo) \Delta M_
{crit}$.  For a high field white dwarf in an MCV
 ($\bo\sim 10^{8}$~G and $\bff\sim 10^{3}$~G), we expect the above
approximate solution
for $\Delta M < \Delta M_b  \sim 10^{-3} \ms$.

\item{\bf Intermediate values of the accreted mass ($\mathbf{\dm
\sim  0.01\ms}$)}

If the accreted mass satisfies the condition $\Delta M_b \ll \dm \ll
\Delta M_{crit}$, then we have the following approximation from
equation (\ref{bt}).

\be
B = 0.8\bff \left(\frac{ \Delta M_{crit}}{\dm}\right)^{7/4}\,.
\label{bdm2}
\ee
Therefore, equation (\ref{bdm2}) implies that the influence of the
initial magnetic field on the magnetic evolution has little effect at
this stage, while the WD magnetic field is scaled by its bottom field
$\bff$.

\item{\bf High values  of the accreted mass ($\mathbf{\dm \sim 0.1 \ms}
$)}

The bottom field $\bff$ is reached when the accreted mass approaches
{\bf $\dm_{f}$} given by
\be \dm_f = 3.5 {} \log(C) \Delta M_{crit} \simeq 0.25 (\ms)
\left(\frac{ \Delta M_{crit}}{0.1\ms}\right)\; \ee Further accretion
has no effect on the value of this field.

\end{itemize}

\begin{figure}
\includegraphics[width=8.0cm, angle=0]{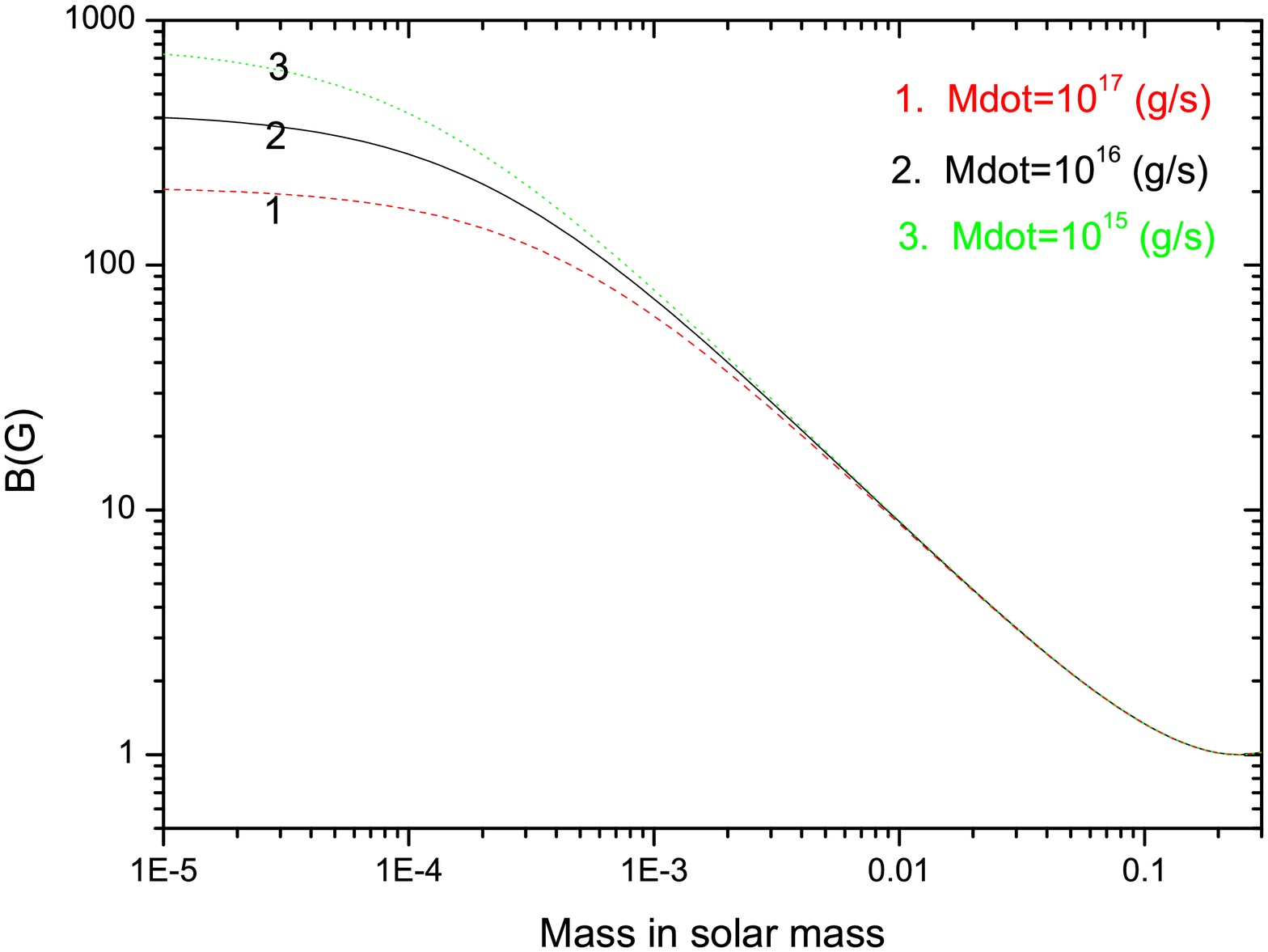}
\caption{The magnetic field versus accreted mass diagram with various
  initial field strengths from $\bo = 10^{6}$ G to $\bo = 10^{8}$ G as
  labeled in the figure, where the mass accretion rate is set to be
  $10^{16}g/cm^{3}$.}
\label{fig-b-dm-bo}
\end{figure}

To summarise, the results of calculations of the evolution of the WD
magnetic field with accreted mass are plotted in Figure
\ref{fig-b-dm-bo}. These show that the polar field decreases as the
accreted mass increases, and saturates once $\sim 0.1\ms$ has been
accreted and a bottom field of $3\times10^{3}$ G has been reached. The
solution is influenced by the initial field only when the total
accreted mass is less than $\sim 0.01\ms$. Beyond this point, the
final (bottom) field depends almost exclusively on the accreted mass,
and the bottom field value of $3\times10^{3}$ G is independent of the
initial field.

\noindent

ZK06 have argued that as the bottom field is reached, the stellar
field lines are redistributed such that, before becoming buried, the
surface equatorial field will approach $B_{\rm eq} \longrightarrow B +
(R/H) (B_{\rm 0} - B)$ and the polar field will reduce to $B\ll B_{\rm
  0}$ yielding $B_{\rm eq} \longrightarrow (R/H) B_{\rm 0} \sim 10
B_{\rm 0}$ if we take $H \sim 0.1 R$.  This means that in an MCV
where the field has been significantly affected by accretion,
a complicated field structure may be expected.

\subsection{Is there evidence for field  {\bf evolution}  in MCVs?}

In order to assess whether the above solutions could be applicable to
the MCVs, we must consider {\bf the} current ideas on the orbital
evolution of CVs and MCVs.  CVs are envisioned to evolve to lower
orbital periods following first contact (initial Roche lobe overflow
after common envelope evolution).  The loss of orbital angular
momentum is driven either by magnetic braking (MB) by the wind that
emanates from the secondary, and/or by gravitational radiation (GR).
A typical system that first comes into contact (say with $P_{\rm orb}
\sim 8$~hr) will evolve to $P_{\rm orb} \sim 3$~hr driven mainly by
MB, at which point the secondary star becomes fully
convective. According to the canonical model, MB is drastically
reduced at this point, and the star, which is by now out of thermal
equilibrium, shrinks within its Roche lobe and mass transfer
ceases. \footnote{The reduction in MB cannot be attributed to the loss
  of magnetic field as had originally been envisaged, since recent
  observations have shown that fully convective M stars also host
  large scale magnetic fields (Morin et al. 2008). The reasons remain
  unclear.}.  The orbit continues to shrink due to the loss of angular
momentum by GR. Mass transfer re-commences when $P_{\rm orb} \sim
2$~hr and the system continues to evolve to shorter periods until the
period minimum is reached (see Warner 1995 for a review). Thus, the
period range $P_{\rm orb} \sim 2-3$~hr, in which there is an apparent
lack of CVs, is referred to as the ``CV period gap''. The evidence for
a period gap is not as strong in the polars (e.g. Webbink \&
Wickramasinghe 2002 and references therein).

As a first approximation, one can assume that the secular evolution of
a CV is driven by a mean mass transfer rate which is determined by the
loss of angular momentum by MB and GR. The mass transfer rate will
then depend mainly on the orbital period of the system. However, there
are significant uncertainties in formulating a MB law that is
applicable to secondary stars of the masses found in CVs. Theoretical
studies of the orbital evolution of CVs have shown that mass transfer
rates of $\sim 7-12 \times10^{16}$ (g/s) are required at the upper end
of the period gap if the observed width of the period gap is to be
reproduced (e.g. McDermott \& Taam 1989; Hameury, King \& Lasota
1991). From the different MB laws that have been considered that
satisfy this requirement, we estimate that a CV that first
comes into contact with a {\bf secondary mass  $M_2=0.5\ms$ and a
primary mass
 $M_1=0.7\ms$} ($P_{\rm
  orb} \sim 5$~hr), $\dot M_{\rm MB} \sim \mbox{a few} \times 10^{17}$
g~s$^{-1}$. On the other hand, $\dot M_{\rm GR} \sim \mbox{a few}
\times
10^{15}$ g~s$^{-1}$ for typical systems in the period range $P_{\rm
  orb} \sim 3-8$~hr.  Also relevant to the following discussion is the
time scale to cross the $2-3$~hr period gap without mass transfer
which can be estimated to be $t_{\rm gap} \sim 4\times 10^8$~yr.

The lack of a pronounced period gap in polars has led to the
suggestion that these systems evolve much more slowly due to a
drastic reduction in the rate of MB.  The reduction is attributed to
the trapping of the stellar wind in the combined magnetosphere of
the two stars in {\bf synchronized } systems (Li \& Wickramasinghe
1998). As a consequence, the secondary remains in near thermal
equlibrium during its orbital evolution, and when the field is
strong enough to severely curtail MB, no period gap is formed (see
Webbink \& Wickramasinghe 2002 for results of detailed
calculations). As a good first approximation, we can assume that the
orbital evolution of most polars is driven by angular momentum loss
at a rate that is closer to the GR rate than to the standard MB
rate.  The mass transfer rate due to GR is typically of a few $\sim
10^{15}$ g~s$^{-1}$ at $P_{\rm orb} \sim 3$~hr which more than an
order of magnitude smaller than the MB rate at the same period.

A {\bf disc} system such as an IP, will transfer matter at a rate
$\sim 10^{17}$~g~s$^{-1}$, that is well
{\ bf within} the regime where an
advective solution may be appropriate. Such a system will {\bf
accrete}  a mass $\Delta M \gsimeq 0.2 \ms$ by the time it reaches
the period gap, if its initial period of contact is $\gsimeq 5$~hr
and mass transfer is conservative.  Thus, a bottom-field state could
be reached.  On the other hand, a typical polar will accrete at a
reduced rate of $10^{15}$ g~s$^{-1}$ and no field evolution or {\bf
decrease in polar field strength} will be expected, even though it may eventually accumulate
a similar amount of mass.

However, it is by no means clear whether {\bf  mass will accumulate
}on the white dwarf during the secular evolution of a CV at the mass
transfer rates estimated above. The orbital evolution is expected to
be punctuated by nova explosions on time scales of $\sim 10^4
-10^5$~yr during which possibly most of the accreted mass will be
ejected from the white dwarf (Townsley \& Bildsten 2004). If this is
the case, the bottom-field state may not be achieved even in systems
that accrete via a disc at a high rate.

In cases where significant field {\bf restructuring}  has occurred, we
expect to observe MWDs in CVs with enhanced equatorial field
strengths relative to what is expected for a centred dipole
configuration. This may be reflected in the presence of dominant
higher order moments in the observed surface field distributions.
{\bf Detailed observations of white dwarfs in MCVs  have enabled
field structures to be determined only  for a handful of
systems}. Although there is evidence for non-dipolar field
structures in many polars, and for the presence of dominant
multipolar components in some cases (e.g. Beuermann et al. 2007;
Reinsch et al. 2005), such large field variations across the stellar
surface are clearly not observed. Thus, if accretion-induced field
{\bf restructuring}  has occurred in polars, it must be at a very modest
level. Unfortunately, we have no knowledge of field structure in
{\bf disc }systems, including the IPs where we may expect the
largest re-structuring/{\bf decline}  of the polar field.
\begin{figure}
\includegraphics[width=8.0cm]{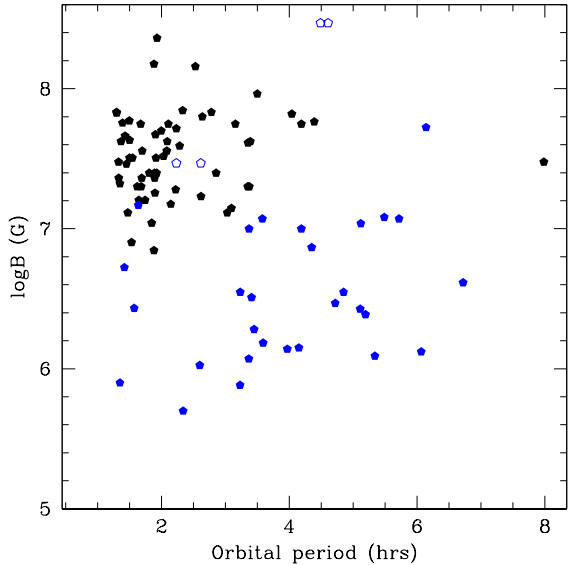}
\caption{Magnetic field versus orbital period diagram for the polars
  (filled circles) and IPs (filled triangles). The empty triangles
  represent IPs which are very close to synchronism. Note that
  the IP magnetic fields are not directly measured, but simply
  inferred (values from Norton et al. 2004).}
\label{Po_B}
\end{figure}

We show in Figure \ref{Po_B} the observed distribution of magnetic
field strengths in MCVs as a function of orbital period.  If we
focus on polars, and assume that they typically evolve from long
orbital periods through the period gap as MCVs, there is no evidence
for an evolution  from high fields to low fields. Indeed, if
at all, the evidence appears to be in the opposite direction.

If we look at the entire class of MCVs (IPs and Polars), we note that
the ratio of the number of polars to IPs is much higher below the gap
as compared to above the gap. This has often been interpreted as
evidence in support of the hypothesis that IPs above the period gap
may evolve into AM Hers below the period gap.  It is indeed the case
that the time-scale to cross the period gap is similar to the
diffusion time scale $t_{\rm diff}(eq)$ for an accreted mass $\Delta M
\sim \Delta M_{\rm crit} \sim 0.2\ms$ (see section
\ref{ohmic_diffusion}). This suggests that a submerged equatorial
field in a system that has evolved as an IP above the period gap may
diffuse outwards during the period gap when accretion ceases. For
$\Delta M \ll \Delta M_{\rm crit}$, as a first approximation, we may
simply assume that accretion screens the field (Cumming 2005). We may
then expect the field to re-emerge to near its original value in the
period gap.  On the other hand, when $\Delta M \sim\Delta M_{\rm
  crit}$, the screening approximation will no longer be appropriate,
and we may expect the field to re-emerge with a significantly {\bf
decreased}  polar field strength. The apparent increase in the ratio of AM Hers to
IPs as one crosses the period gap could partly be due to screening
above the period gap. However, the dominant effect may be the
greater ease with which systems can lock into synchronism below the
gap due to the combination of a lower mass transfer rates and a
shorter orbital period (Wu \& Wickramasnghe 1991).

It is possible that systems that evolve from sufficiently long
orbital periods may accrete sufficient mass even in the presence of
nova eruptions to reach a bottom-field state. Therefore, this
depends on the efficiency with which mass is ejected in a nova
explosion which remains uncertain. We predict that if such systems
exist, they will form a sub-class of CVs {\bf with discs} with distinctive
homogeneous properties. Their polar field strengths would be
expected to be $B_{\rm f} \sim 3\times 10^3$~G and they should have
strongly non-dipolar magnetic field structures.

\begin{figure}
\includegraphics[width=8.0cm,angle=0]{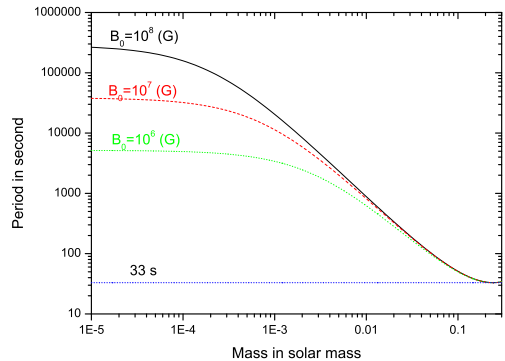}
 \caption{The spin period of WD versus accreted mass diagram with
various initial field
conditions, where the mass accretion rate is set to be
$10^{16}$~g~cm$^{-3}$.}
\label{fig-p-dm-bo}
\end{figure}

Their spin frequency should be a fraction $\psi$ of the Keplerian
spin frequency at the magnetospheric radius $R_{\rm M}\sim R$: \be
\ns = 0.03 (Hz) \left(\frac{\psi}{0.5}\right)~A ,\qquad P = 33~({\rm
s}) \left(\frac{0.5}{\psi}\right)\left(\frac{1}{A}\right) \ee where
\be A=\left(\frac{M}{\ms}\right)^{1/2}\left(\frac{R}{10^{9}{\rm
    cm}}\right)^{-3/2}\nn \ee Observations of Low Mass X-ray Binaries
(LMXBs) have shown that the maximum spin frequency of a neutron star
is 619 Hz, which is roughly half the measured maximum kHz QPO
frequency 1330 Hz, that is conventionally ascribed to the Keplerian
frequency (van der Klis 2000; 2006).  By analogy, we may assume
$\psi \sim 0.5$ which suggests that systems that reach the
bottom-field state should have spin periods of $\sim 30$~s which is
close to the fastest spin periods observed in CVs. We show in Figure
\ref{fig-p-dm-bo} the relation between accreted mass and spin period
for different values of the initial magnetic field $B_0$ at a fixed
mass accretion rate of $10^{16}$~g~s$^{-1}$.

Magnetic fields are expected to play a role in explaining QPO
behaviour in the accreting neutron stars in Low Mass X-ray Binaries
(LMXBs). In many respects, the properties of the QPOs observed in the
dwarf novae parallel what is seen in neutron stars in the LMXBs (see
the review by Warner \& Woudt 2008 and references therein) and so it
may be tempting to identify the dwarf novae (DN) that exhibit QPO
oscillations with MCVs that have reached the bottom-field state.

It should be noted that only $25$\% of CVs are magnetic and, if at
all, only a small fraction of these systems could have reached the
bottom-field state. Thus, one should expect that the vast majority of
the DN are likely to be systems that are born with low fields
($\lsimeq 10^3$ ~G) and a distribution similar to that inferred for
the low field isolated magnetic white dwarfs (see next section).

\subsection{The field distributions of MCVs and isolated magnetic
white dwarfs}

As noted in the introduction and illustrated in Figure
\ref{magCVs_fields}, there are distinct differences in the field
distribution of MCVs and of isolated HFMWDs.  The peak of the
distribution for MCVs occurs at $3\times 10^7$ compared to $\sim
10^7$~G for the HFMWDs.  Generally, compared to the HFMWDs, there is a
dearth of low field systems and the very high field systems are not as
well represented among the MCVs.  The lack of high field systems may
be simply related to the extreme reduction that is expected in MB, and
hence in the mass transfer rate in such systems which will make them
faint and thus much more difficult to detect (e.g. Li, Wu \&
Wickramasinghe 1994). The apparent lack of low-field MCVs, however, is
harder to explain.

\begin{figure}
\includegraphics[width=8.0cm,angle=0]{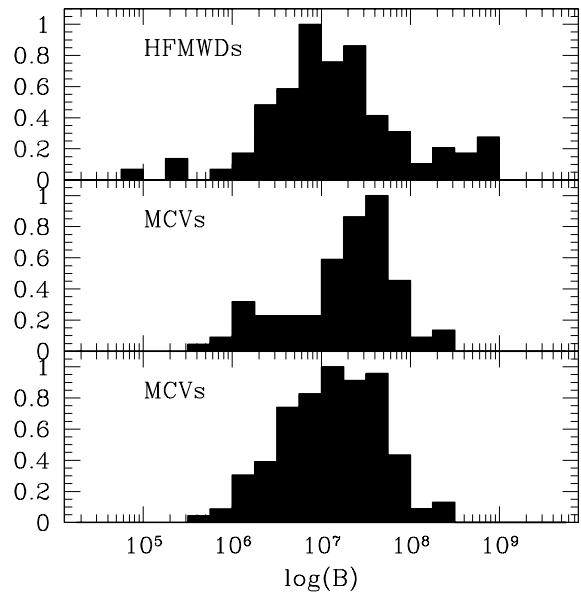}
 \caption{Top panel: the magnetic field distribution of the
   HFMWDS. Middle panel: the magnetic field distribution of the
   MCVs. Here the field strength of most IPs has been inferred on the
   assumption that the white dwarf is in spin equilibrium with an
   accretion disc. Bottom panel: the magnetic field distribution of
   the MCVs obtained by assuming that 40\% of polars with no field
   determinations have magnetic fields between $\sim 2-20$~MG (see
   text).}
\label{mcv_mwds_fields}
\end{figure}

In assessing the statistics, we note that there are currently about
40\% of polars with no field determinations.  It is possible that
observations will reveal that most of these systems have magnetic
fields between $\sim 2-20$~MG.  If we assume that this is the case,
and that the fields are distributed as a Gaussian in the logarithm
with a mean $\langle \log\rangle B=6.8$ and a standard deviation
$\sigma_{\log B}=0.25$, we obtain the results shown in the bottom
panel of Figure \ref{mcv_mwds_fields}. The distribution of magnetic
fields in MCVs and HFMWDs then become similar, and a strong case
cannot be made for field decay.  On this scenario, it has to be
assumed that most systems that are observed as IPs above the period
gap will synchronize in the period gap and be seen as polars below
the gap thus explaining the observed reduction in the ratio of IPs
to polars below the gap.

On the other hand, if observations show that the 40\% of polars with
no field determinations have fields similar to those measured in the
remaining 60\% of polars, then the reduction in the ratio of IPs to
polars below the gap can be attributed {\bf to effects of accretion on the
field}.

The magnetic properties of the isolated LFMWDs and their incidence
are less well established, {\bf due to }current limitations of
measuring fields $\lsimeq 10^3$~G. Thus, it is possible that either
all WDs are magnetic at some level or that most WDs are essentially
non magnetic. In any case, the LFMWDs are likely to be associated
with the vast majority of the CVs that show no direct evidence for
coherent pulsations at the spin period of the white dwarf as would
be expected if there were magnetically channelled accretion on to
the surface of the white dwarf . Unfortunately, we have no direct
evidence of the field strengths of the white dwarfs in such {\bf CVs
with discs} (e.g. in dwarf novae).

\section{Summary and conclusions}\label{conclusions}

The anomalously low fields that have been deduced for the neutron
stars in the majority of LMXBs ($10^8 -10^{10}$~G) compared to the
magnetic fields in the isolated radio pulsars ($10^{11} -10^{14}$~G)
is seen as strong evidence for accretion-induced field decay
(e.g. Manchester 2006). In these disc accretors, the competition
between the advection of the field from polar to equatorial regions,
and the tendency of the field to re-emerge by Ohmic diffusion
results in a net reduction of the field at least during phases of
accretion. On the other hand, if the LMXBs do evolve into the BMSPs
as is generally assumed (but see Hurley et al. 2009; Ferrario \&
Wickramasinghe 2007), one can also conclude that the field does not
re-emerge when accretion ceases, {\bf  at least} not on the time
scale of the typical age ($\sim \mbox{a few} 10^9$~yr) of a BMSP.
The latter conclusion would imply that accretion does not simply
lead to a screening of the magnetic field  which subsequently
re-emerges to near its original value, but would yield a change in
the current forming regions, and thus to field decay.

We have investigated whether a similar process may be in operation in
accreting MWDs in the CVs. In particular, we have asked whether there
is evidence for the existence of group of systems that have reached a
bottom field among the CVs.

We have argued that the mass transfer rates that determine the secular
evolution of CVs is different for systems that evolve as polars or as
IPs. Based on our simplified model for the evolution of the field due
to advection, and estimates of times scales of Ohmic diffusion, we
conclude as follows.

\begin{enumerate}
\item In the advection dominated regime, we expect that as matter
  accretes on to the polar caps, the field lines will be dragged from
  the polar caps towards the equator. The bottom-field state is
  reached when $\sim 0.2\ms$ is accreted and the equatorial field
  becomes buried below the white dwarf surface.  Systems that enter a
  bottom-field state will have no memory of initial conditions
(initial
  field and spin period) and are expected to form a homogeneous group
  with $B_{\rm f} \sim 3\times10^{3}$ G, and $P \sim 30 $~s.

\item The advection dominated solution is expected to be applicable
  only for accretion rates above a critical value $\dot M_{\rm crit}
  \sim10^{16}g/s$. Below this rate, the field will diffuse outwards
  faster than it is advected. The polars, with typical secular mass
  transfer rates of a few times $10^{15}$~g~s$^{-1}$, are therefore
not expected to
  exhibit significant {\bf field restructuring and reduction in polar field strength}.
  This is consistent with the
  general lack of evidence for field evolution with orbital period in
  polars.

\item The advection dominated solution is expected to be applicable to
  the IPs above the period gap, which typically have accretion rates
  of $\sim 10^{17}$~g~s$^{-1}$. Systems that first come into contact
  at $P_{\rm orb} 5$~hr will accrete $\Delta M 0.2 \ms$ as they
  reach the upper end of the period gap ($P_{\rm orb} \sim 3 $~hr) if
  mass transfer is conservative. However, when nova eruptions are
  taken into consideration, the mass that is built up by accretion may
  be significantly smaller. We may therefore expect only a small
  proportion of IPs, if any, to reach the bottom-field state.

\item If there are MCVs that have reached the bottom-field state, they
  are likely to be found among the dwarf novae that exhibit QPOs. They
  may be expected to have very peculiar field structures dominated by
  strong equatorial field enhancements. However, this prediction
cannot be easily
  verified, since we have little information on the field structures
  of white dwarfs in disc accreting systems.

We conclude by noting that there are two major differences between
LMXBs and CVs that make field  {\bf restructuring and /or decay }
 more likely in the
LMXBs.  Due to vastly different magnetic moments of the neutron
stars (typically lower by a factor $\sim 10^6$), there are no
magnetically phase locked systems similar to the polars. Thus, while
the majority of the CVs that are recognised to be magnetic are the
strong field polars, which we have argued are unlikely to exhibit
{\bf accretion induced field restructuring}  due to their
intrinsically lower mass accretion rates, all the LMXBs (which
include both high and low field neutron stars) are disc systems with
intrinsically higher mass transfer rates, and are therefore more
likely to exhibit field {\bf restructuring and /or decay}. {\bf A
second major difference is that thermonuclear runaways similar to
nova explosions in CVs are not expected to result in mass loss from
neutron stars, because of their deep gravitational potential, so
that the critical mass is likely to be achieved more easily during
LMXB evolution.}

\end{enumerate}

\section*{Acknowledgements}
This research has been supported by NSFC (No.10773017) and National
Basic Research Program of China (2009CB824800).


\end{document}